\documentclass[aps,prb,amssymb,showpacs,floatfix, twocolumn, bibliography]{revtex4-1}
\usepackage{amssymb,graphicx}
\usepackage{epsfig,graphicx}
\usepackage{subcaption}
\usepackage{subfloat}
\usepackage{epstopdf}
\usepackage{amsmath}
\usepackage{tikz}
\usepackage{xr}
\usepackage{float}

\usepackage{hyperref}
\usepackage{cleveref}

\begin{document}
\title{Surface states in Dirac metals and topological crystalline insulators}

\author{Grigory Bednik}

\affiliation{Department of Physics and Astronomy, University of Waterloo, Waterloo, Ontario, N2L 3G1, Canada}

\date{May 30, 2018}
\newcounter{TypeOne}
\newcounter{TypeTwo}

\setcounter{TypeOne}{1}
\setcounter{TypeTwo}{2}

\begin{center}
\begin{abstract}

We reconsider the problem of surface states spectrum in type \Roman{TypeOne} Dirac metals. We find that the surface states, despite being gapped, always form branches terminating at Dirac points. Furthermore, we consider evolution of the surface states in the case, when rotational symmetry is broken, and as a result, Dirac points are gapped.  We find, that in this case, special role is played by mirror symmetry relative to the plane connecting Dirac points. When it is present, the resulting gapped state is a topological crystalline insulator, which surface spectrum can contain either one or three Dirac points, two of which are protected solely by the mirror symmetry. The Dirac metal can be viewed as a topological phase transition between two phases with different mirror Chern numbers.  

\end{abstract}
\end{center}

\maketitle

\section{Introduction}

Topological materials are known to have distinct properties, which can change only through a phase transition. The first discovered example of a topological insulator is a $\mathrm{HgTe}$ quantum well \cite{Bernevig1757, Konig766}, which exhibits topological phase transition between normal and topological phase at critical thickness $d_c = 6.3\mbox{nm}$. At the critical point, energy gap closes, and as a result, a new state with Dirac-like dispersion emerges. This idea was used to create gapless critical phases in $\mathrm{3D}$ topological insulators  $\mathrm{Bi Tl (S_{1-\delta} Se_{\delta})_2 }$ \cite{Xu560},   $({\mathrm{Bi}}_{1\ensuremath{-}x}{\mathrm{In}}_{x}{)}_{2}{\mathrm{Se}}_{3}$ \cite{PhysRevLett.109.186403}, topological crystalline insulator $\mathrm{Pb_{1-x}Sn_x Te}$ \cite{NatCommun3.2191} and other materials. 

 The concept of topological materials has expanded to topological metals. The simplest example of a topological metal is Weyl metal \cite{PhysRevB.93.195138, PhysRevB.83.205101, PhysRevLett.107.127205, NatCommun6.8373, NatPhys10.3425} - its two non-degenerate bands touch at isolated pairs of Weyl points, each of which is a source or a sink of Berry curvature. Weyl metals are known for their Fermi arcs - surface states connecting the Weyl points. Their existence is protected topologically - they exist at such points in momentum space, where a plane passing between the Weyl points in a Brillouin zone has non-zero Chern number.

In the recent years, the notion of topological metals was generalized to Dirac metals  \cite{PhysRevLett.108.140405, PhysRevB.85.195320, Science343.864, NComms5.5898, PhysRevLett.113.027603, PhysRevB.88.125427, NatCommun5.4786,  NatMatt13.677}: materials with bands touching at Dirac points, near which the dispersion is described by four-component Dirac equation. In contrast to the Weyl metals, Dirac points are protected not topologically, but by discrete rotational symmetry: bands touching at Dirac point cannot be gapped because they are eigenstates of rotation operator with different eigenvalues. This implies that, by breaking the rotational symmetry, Dirac metal can be turned into an insulator with different possible properties \cite{PhysRevB.93.045118, PhysRevB.91.214517, SciRep6.24137, PhysRevB.96.075112, C4CP05115G}. The fact that Dirac points are not topological, affects the properties of surface states: in contrast to Weyl metal,  Fermi arcs in Dirac material are, generally, gapped \cite{NatCommun5.6161, Kargarian02082016}.

The idea of Dirac metal as a critical point between two insulating phases was proposed in the Ref. \cite{PhysRevB.78.045426} (in Ref.\cite{PhysRevB.90.245308} there was also proposed a phase transition between stable Dirac metal and insulator phases). In particular, it was suggested, that in an alloy $\mathrm{Bi_{1-x} Sb_{x}}$, Dirac metal emerges as a critical point between normal and $\mathbb{Z}_2$ topological phases at $x \sim 0.03$. Furthermore, it was suggested that such critical point is characterized not only by change of $\mathbb{Z}_2$ invariant, but also by change of mirror Chern number. This fact, in turn, must manifest itself in the surface states: the surface Dirac cones of topological insulators with opposite mirror Chern numbers have opposite helicities  \cite{PhysRevB.91.165435}.

All known Dirac metals can be classified into two categories based on the location of their Dirac points: they can be located either at arbitrary points along rotation axis (type \Roman{TypeOne}), or at time-reversal invariant momenta (type \Roman{TypeTwo} ) \cite{NComms5.5898}. Two experimentally discovered examples of the former type are $\mathrm{Na_3 Bi}$ \cite{PhysRevB.85.195320, Science343.864} and $\mathrm{Cd_3 As_2}$ \cite{PhysRevB.88.125427, NatCommun5.4786, NatMatt13.677}, and an example of the latter is $\mathrm{ZrTe_5}$ \cite{PhysRevLett.115.176404}. However, all known models of Dirac metals as critical points between different topological phases \cite{Xu560, PhysRevLett.109.186403, PhysRevB.78.045426}, possess a single Dirac point located at the center of the Brillouin zone. It has not yet been demonstrated that Dirac metal with two spatially separated Dirac points can also be a critical point between two different topological phases.

In this work, we consider general model of type \Roman{TypeOne} Dirac metal and explore the form of its surface spectrum. Our findings confirm previous results \cite{Kargarian02082016}, saying that surface states are generally gapped except the special plane $k_z = 0$. However, we claim that, despite being non-topological, surface states in Dirac metal still terminate at Dirac points.  Cross-section of the surface states with Fermi level, i.e. Fermi arcs consist of three surfaces: one ring surrounding  point with time-reversal invariant momentum, i.e. $\vec{k}=0$ (which was found in the Ref. \cite{Kargarian02082016}), and two arcs 'attached' to the bulk Fermi surface.

Furthermore, we explore the surface states spectrum in the case, when rotational symmetry protecting the bulk Dirac points is broken. We find that special role is played by mirror symmetry relative to plane connecting the bulk Dirac points. Specifically, we find that in the presence of such symmetry, the resulting phase is mirror Chern insulator, and the Dirac metal is a critical point, where mirror Chern number changes its sign. This fact affects the surface states: in the mirror invariant plane, they can be characterized by their mirror eignvalue, and states with its fixed value change their 'handedness' at different sides of the transition. This, in turn can be used to illustrate that in the critical point, i.e. in the Dirac metal, surface states always terminate at the bulk Dirac points. In the insulating phases on both sides from the transition, the surface states have either one or three Dirac points, two of which are protected solely by the mirror symmetry: if the latter is broken, two surface Dirac points become gapped. 

This paper is organized as follows. In Sec. \ref{OurModel} we present our model of Dirac metal, and demonstrate explicitly that, by breaking rotational symmetry, it can be converted into topological crystalline insulator. In Sec. \ref{SurfaceStates}, we describe their surface states. Specifically, in part \ref{SurfaceStatesDiracMetal} we describe surface states structure in Dirac metal, and in part \ref{SurfaceStatesTCI}, we describe the surface states structure of TCI phase. In part. \ref{BrokenMirrorSymmetry}, we discuss the effect on surface states from breaking of the mirror symmetry. We summarize our findings in Sec. \ref{Conclusions}.


\section{Model of Dirac metal and topological crystalline insulator}
\label{OurModel}
\begin{figure}
	\includegraphics[width=4.5cm,angle=0] 
	{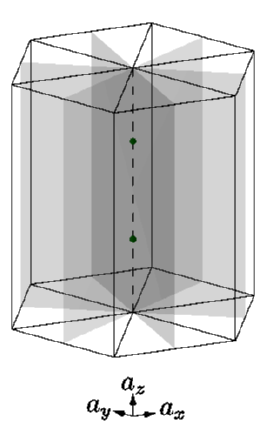}
\caption{A schematic plot of the Brillouin zone in the Dirac metal $\mathrm{Na_3 Bi}$. The Dirac points are aligned along the $z$ direction. Three mirror planes along the line connecting them are shown. Below the figure, Bravais lattice vectors (in coordinate representation) are drawn to demonstrate their direction relative to the mirror planes. }
\label{BZ}
\end{figure}

As it was shown in \cite{NComms5.5898}, Dirac metals require the presence of time-reversal, inversion and rotational symmetries. The first two of them are necessary to ensure double degeneracy of the bands, and the latter is necessary to stabilize the Dirac points. Dirac metals with spatially separated Dirac points can be classified by their rotational symmetry properties (three-, four-, and six-fold), and by their inversion symmetry matrix (unity or $\sigma_z$). Experimentally discovered Dirac metal $\mathrm{Na_3 Bi}$  (its crystal group is $P6_3/mmc$) belongs to symmetry group containing three-fold rotations, and it is described by a model with the inversion matrix $\sigma_z$. However, the full crystal group of $\mathrm{Na_3 Bi}$, in addition, possesses mirror symmetries over the planes $(x, 2x, z)$ , $(2x, x, z)$, and $(x , \bar{x} , z)$, i.e. the planes lying along the line connecting the Dirac points (see Fig. \ref{BZ}). We would like to explore their effect in more details. 

Since we are interested only in general features of our low energy model, we expect that the experimentally relevant case of three-fold rotational invariance is not qualitatively different from the case of four-fold rotational invariance. Thus, for simplicity, we consider lattice Hamiltonian invariant under four-fold rotations. We write it as \cite{PhysRevB.85.195320, NComms5.5898, PhysRevLett.117.136602}:
\begin{eqnarray}
H_0 &=& v_F\sin k_x \Gamma_1 + v_F  \sin k_y \Gamma_2 + m(k) \Gamma_3
\label{DiracHam}
\\
 &&+ \gamma k_z (\cos k_x -  \cos k_y ) \Gamma_4 + \gamma k_z  \sin k_x  \sin k_y  \Gamma_5,
\nonumber
\end{eqnarray}
and define the gamma-matrices as:
\begin{eqnarray}
\Gamma_1= \sigma_x \otimes s_z, \qquad \Gamma_2 = -\sigma_y, \qquad \Gamma_3 = \sigma_z,
\nonumber\\
\Gamma_4 = \sigma_x \otimes s_x, \qquad \Gamma_5 = \sigma_x \otimes s_y,
\nonumber
\end{eqnarray}
where $\sigma$ refers to orbital, and $s$ - to spin degrees of freedom. $m(k)$ is a function of momentum within the Brillouin zone (BZ), which changes sign at the Dirac points. For simplicity, we take as: $m = m_0 - 2b_{xy}( 2 - \cos k_x - \cos k_y) - b_z (1-\cos k_z)$. We assume that values of the parameters are such that $m(k)$ changes sign along the line $k_x = k_y = 0$, and thus has a pair of Dirac points.

The bands in this model are doubly degenerate due to the presence of both time-reversal and inversion symmetries, which are defined as $T = is_y \mathcal{K}$ (here $\mathcal{K}$ is an operator of complex conjugation) and $P = \Gamma_3$ correspondingly. Dirac points are stabilized by the rotational symmetry around $z$ direction with the generator $C = (2-\sigma_z) \otimes s_z$: states with positive energy have eigenvalues of the rotational symmetry operator $j = \pm1/2$, whereas states with negative energy have rotational eigenvalues $j=\pm 3/2$ (as in $\mathrm{Na_3 Bi}$). 

In addition, our model possesses two mirror symmetries in $xz$ and $yz$ planes, i.e. the planes lying along the Dirac points. The mirror symmetry over $xz$ plane in written as:
\begin{eqnarray}
H(k_x , -k_y , k_z) = \mathcal{M} H(k_x , k_y , k_z) \mathcal{M}^{-1},
\end{eqnarray}
where $\mathcal{M} = i \Gamma_2 \Gamma_5$ is the generator of the mirror symmetry relative to $xz$ plane.

To break the rotational, while preserving the mirror symmetry of the model (\ref{DiracHam}), we add the following term:
\begin{eqnarray}
H_1 = \Delta_1 \sin k_z \Gamma_4.
\label{Gamma4Term}
\end{eqnarray}
In the presence of this term, Dirac points become gapped, i.e. the system becomes a strong topological insulator (one can see this by looking at inversion eigenvalues of the bands at time-reversal invariant momenta). We will see shortly, that this system is actually a topological crystalline insulator. 

The presence of mirror symmetry makes it possible to split the space of all states in the mirror invariant plane (i.e. at $k_y=0$)  into two subspaces according to their mirror eigenvalues $\pm1$. Thus, at $k_y=0$, the total Hamiltonian defined by the Eqs.  (\ref{DiracHam}-\ref{Gamma4Term}) decouples into two $2\times2$ Hamiltonians $H = H_{+}  \oplus H_{-}$, each describing states with mirror eigenvectors $\pm1$. These Hamiltonians have the following explicit expressions:
\begin{eqnarray}
H_{\pm} &=& v_F \sin k_x \tau_1  \pm  m(k) \tau_3
\nonumber\\
 &&+ \left( \gamma (1-\cos k_x)  + \Delta_1 \right)  \sin k_z \tau_2.
\label{HFixedMirror}
\end{eqnarray}

One can find that if $v_F$ is fixed positive, two filled bands of $H_{\pm}$ have Chern numbers $\pm \mathrm{sign} \Delta_1$ (for details of the calculation see \ref{CN}), i.e. the bands with opposite mirror eigenvalues have opposite Chern numbers (CNs). Thus, at $k_y=0$, the gapped system defined by the Eqs. (\ref{DiracHam}-\ref{Gamma4Term}) has zero total CN, but non-zero mirror CN, which makes the whole system be topological crystalline insulator. 

 An important feature here is that the mirror Chern number changes its value from $+1$ to $-1$, as $\Delta_1$ changes sign, i.e. at the critical point, where the Dirac metal is realized. Thus we conclude that Dirac metal phase at $\Delta_1 = 0$ is a topological phase transition between two insulators with opposite mirror Chern numbers.


\section{Surface states}
\label{SurfaceStates}

It is known, that topological properties of a band structure are related to its surface states spectrum, and therefore it is of interest to study it in our model. Since our model possesses mirror symmetry in the $xz$ plane, and thus the total $4 \times 4$ Hamiltonian  (\ref{DiracHam}-\ref{Gamma4Term}) at $k_y=0$ can be decoupled into two $2\times 2$ Hamiltonians (\ref{HFixedMirror}), surface states in the mirror invariant plane can be found analytically in the long wavelength limit. At $k_y \ne 0$, the total Hamiltonian cannot be decomposed in a similar way, and therefore the surface states can be found only numerically. Because of that, in this section we present numerical solutions of the model (\ref{DiracHam}) at arbitrary $k_y$, including the special case $k_y=0$, and in the Sec. \ref{SecSurfaceStates}, we present the analytical calculation of the surface states at $k_y=0$, which demonstrates generality of our result. In part \ref{SurfaceStatesDiracMetal} we describe the structure of surface states in Dirac metal, and in part \ref{SurfaceStatesTCI} we describe their behavior after opening the gap, i.e. in topological crystalline insulators. Throughout our analysis, we assume that the surface is located at the plane $x=0$.


\subsection{Surface states in Dirac metal}
\label{SurfaceStatesDiracMetal}

As it was established previously, Dirac metal, similarly to topological insulator, possesses $\mathtt{Z}_2$ invariant, and therefore, its surface states contain Dirac cone with the center at $k_y = k_z = 0$. In the special case of conserved spin, i.e. at $\gamma=0$, Dirac metal (\ref{DiracHam})  is formed by two copies of Weyl metal, and therefore its surface states have the same structure, as two copies of surface states of Weyl metal. More specifically, at $k_y=0$ and any $k_z$ between the bulk Dirac points, the spectrum is gapless and doubly degenerate. In other words, the surface states form Fermi arcs connecting the bulk Dirac points. However, once $\gamma$ is turned on, the surface spectrum at any $k_z \ne 0$ becomes gapped. Fermi arcs, i.e. crossings of the surface bands with the Fermi level do not connect the Dirac points anymore. 

As we mentioned previously, the presence of mirror symmetry in the $xz$ plane, i.e. the plane containing the Dirac points, makes it possible to find the surface states at $k_y=0$ analytically. After computing them explicitly, we find that despite being non-topological, surface states in Dirac metal always terminate in Dirac points, where their energy reaches zero and penetration depth becomes infinite. We present their structure at $k_y=0$ on the Fig. \ref{Delta_0}. This fact makes the surface states in Dirac metal similar to surface states in two copies of Weyl metal.

\begin{figure}[H]
\centering
	\begin{subfigure}[t]{0.24\textwidth}
		\includegraphics[width=4.7cm,angle=0] 
		{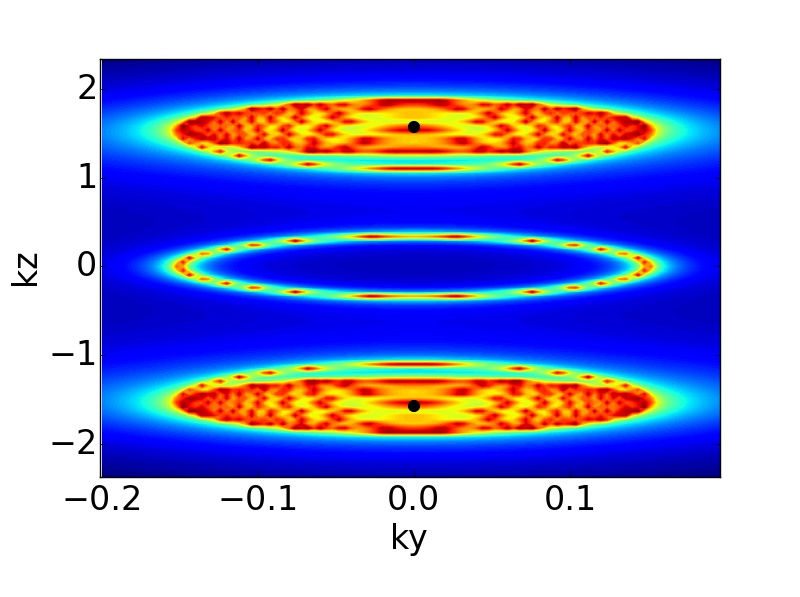}
		\vspace{-0.5cm}
		\subcaption{}	
		\label{E015}
	\end{subfigure}
	\begin{subfigure}[t]{0.23\textwidth}
		\includegraphics[width=4.7cm,angle=0]
		{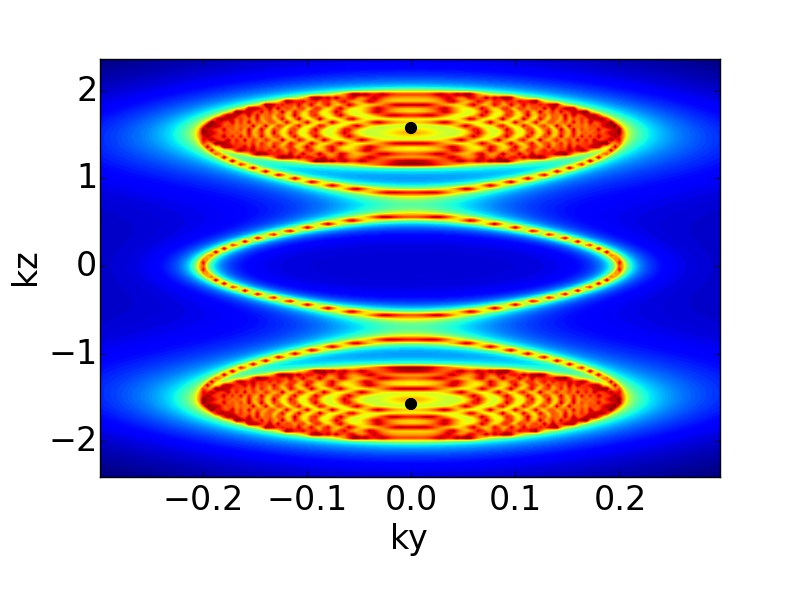}
		\vspace{-0.5cm}
		\subcaption{}
		\label{E020}
	\end{subfigure}
	\begin{subfigure}[t]{0.23\textwidth}
		\includegraphics[width=4.7cm,angle=0]
		{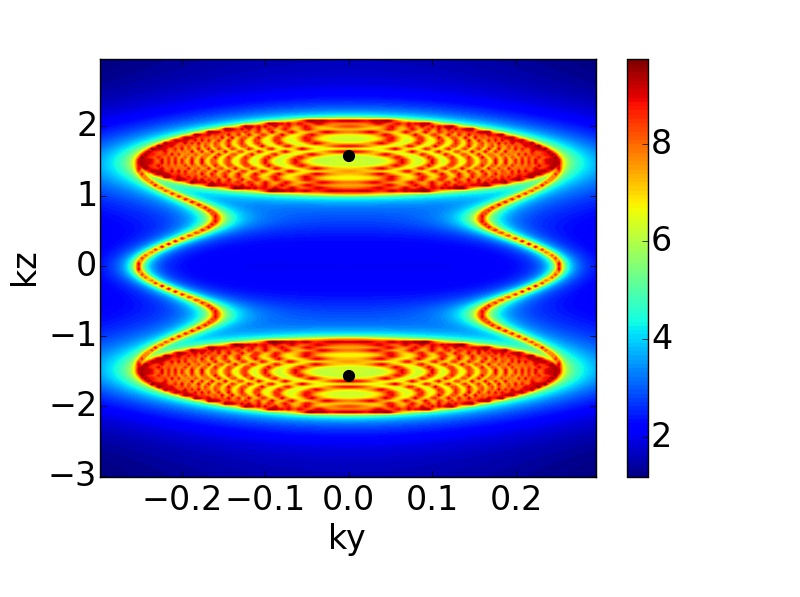}
		\vspace{-0.5cm}
		\subcaption{}
		\label{E025}
	\end{subfigure}

\caption{Spectral plots, showing the structure of states in the model of Dirac metal (\ref{DiracHam}) at fixed Fermi levels: (\ref{E015}) $0.15$, (\ref{E020}) $0.20$, (\ref{E025}) $0.25$. The parameters are: $v_F=1.0$, $m=0.5$, $b_{xy}=0.5$, $b_z = 0.5$, $\gamma=1.0$. The position of Dirac points is shown by black dots.
}
\label{FermiArcs}
\end{figure}

Outside the mirror plane, i.e. at $k_y \ne 0$, the surface states can be found only numerically.  They form two symmetrical 'gorges' with the bottoms at $k_y=0$. As we said before, they are degenerate either at the center of the Brillouin zone (i.e. at $k_z=0$), or at the bulk Dirac points. On the Fig. \ref{FermiArcs} we present the structure of the surface states crossings with the Fermi level. At small Fermi energy, the surface states form several Fermi lines: one closed line created from the surface Dirac cone at the BZ center, and two open lines terminating at the bulk Dirac cones (see Fig \ref{E015}). As the Fermi level increases, the surface Fermi lines also expand and eventually merge with each other (Fig. \ref{E020}). Thus, at sufficiently large Fermi level, there exist two Fermi arcs connecting the opposite bulk Dirac cones (Fig. \ref{E025}).

Finally, we note that (as it is evident from the analytical solution Eq. \ref{EnergyAnswer}) the described above structure of the surface states persist at any value of $\gamma$: as it increases, the gap in the surface spectrum also increases, and at infinite $\gamma$  the surface spectrum merge with the bulk spectrum.


\subsection{Surface state in the TCI}
\label{SurfaceStatesTCI}

In this section, we study the case, when the model of Dirac metal (Eq. \ref{DiracHam}) acquires a gap due to the term, breaking the rotational symmetry (\ref{Gamma4Term}). As we mentioned, such model still possesses mirror symmetry in the $xz$ plane, and, as a result, its Hamiltonian at $k_y=0$ decouples into two Hamiltonians (\ref{HFixedMirror}) describing states with fixed mirror eigenvalues. Thus we can consider each sector with the fixed mirror eigenvalue separately. 

\begin{figure}[h]
\centering
	\begin{subfigure}[b]{0.24\textwidth}	
		\includegraphics[width=4.8cm,angle=0] 
		{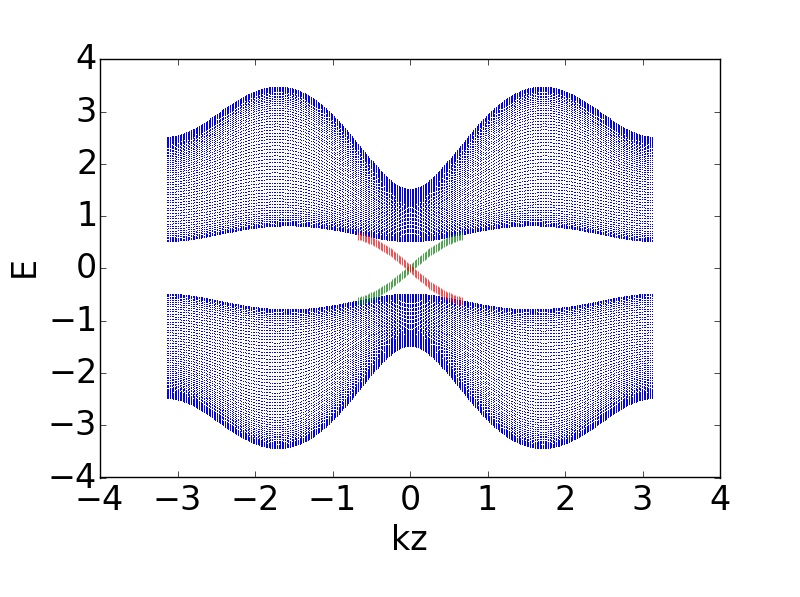}
		\vspace{-0.5cm}
		\subcaption{}	
		\label{Delta_08}
	\end{subfigure}
	\begin{subfigure}[b]{0.23\textwidth}	
		\includegraphics[width=4.8cm,angle=0]   
		{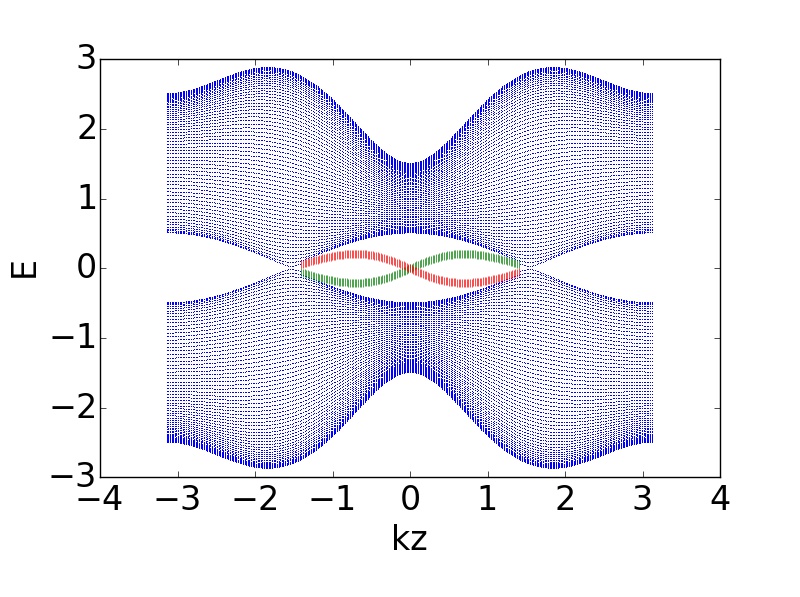}
		\vspace{-0.5cm}
		\subcaption{}
		\label{Delta_0}
	\end{subfigure}
	\begin{subfigure}[b]{0.24\textwidth}	
		\includegraphics[width=4.7cm,angle=0]
		{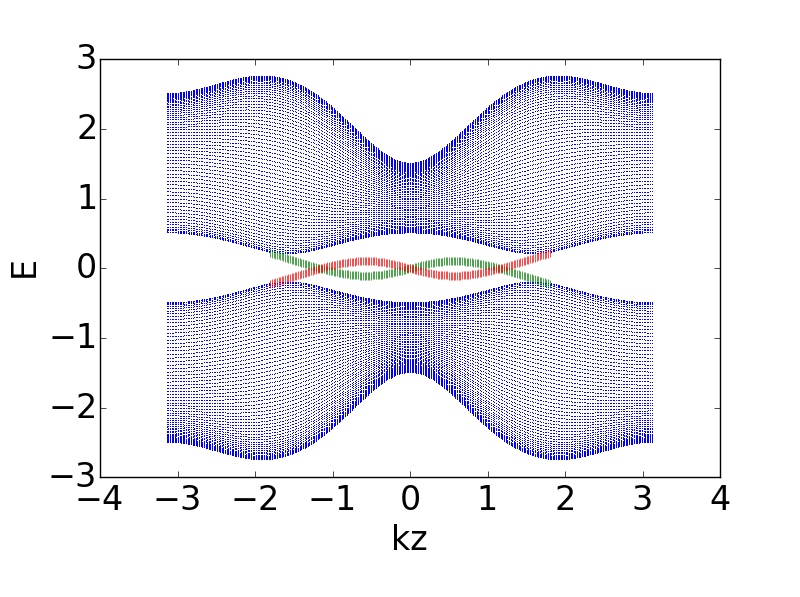}
		\vspace{-0.5cm}
		\subcaption{}
		\label{Delta_m02}
	\end{subfigure}
	\begin{subfigure}[b]{0.23\textwidth}	
		\includegraphics[width=4.7cm,angle=0]
		{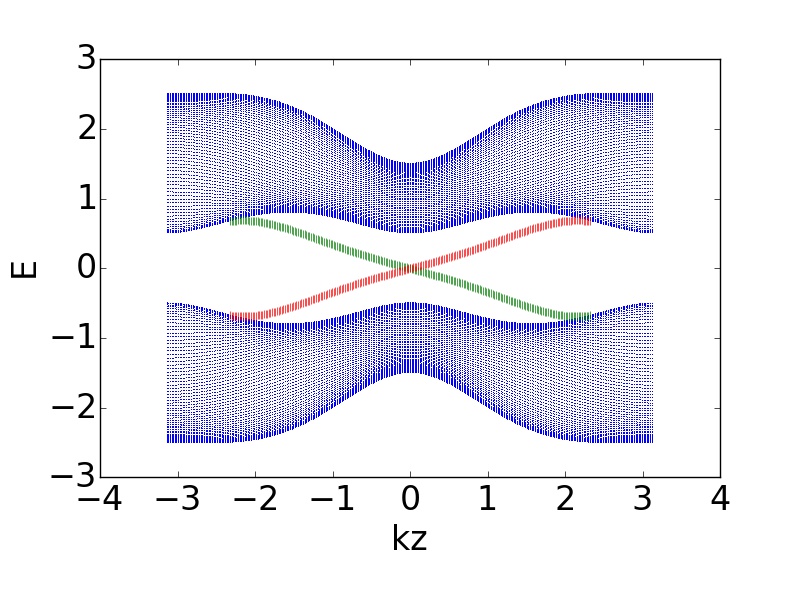}
		\vspace{-0.5cm}
		\subcaption{}
		\label{Delta_m08}
	\end{subfigure}
\caption{Dispersion strucure of the model (\ref{HFixedMirror}) and mirror eigenvalue $+1$. The values of $\Delta_1$ are $0.8$ (\subref{Delta_08}), $0.0$ (\subref{Delta_0}), $-0.2$ (\subref{Delta_m02}), $-0.8$ (\subref{Delta_m08}). The parameters are: $v_F=1.0$, $m=0.5$, $b_{xy}=0.5$, $b_z = 0.5$, $\gamma=1.0$. Blue color refers to bulk states; Red and Green colors refer to surface states localized at the left and right boundary correspondingly. Surface states with a given mirror eigenvalue and localized at one surface have the same dispersion as states with the opposite mirror eigenvalue and localized at the opposite surface.  }
\label{SurfaceStates}
\end{figure}

Let us focus on the case of the model (\ref{HFixedMirror}) with \textit{positive} sign. The case of \textit{negative} sign can be obtained from the former by appropriate transformation. We have established that the model $H_{+}$ has Chern number $\pm 1$ at positive/negative $(\Delta_1)$, and therefore, it is expected to possess surface states propagating in the positive/negative direction. This is indeed confirmed by both numerical and analytical calculation (see Fig. \ref{SurfaceStates} ). At $\Delta_1 >0$ there exists an insulating phase with the right-handed surface states (Fig. \ref{Delta_08}). As $\Delta_1$ approaches zero, the gap closes and thus the phase transition occurs. As $\Delta_1$ becomes negative, the Chern number becomes equal to $-1$, and thus the surface states propagate in the negative direction (see Figs. \ref{Delta_m02} , \ref{Delta_m08}). 

 In the special case of conserved spin, i.e. at $\gamma=0$, the change in the direction of propagation occurs through flat band (in such case, surface states in Dirac metal form gapless Fermi arcs). However, at arbitrary $\gamma$, the surface Dirac point at $k_z=0$ still possesses finite Fermi velocity at the critical point (see Fig. \ref{Delta_0}), and thus the change of 'handedness' occurs through the formation of two new Dirac points in the surface spectrum (see Fig.\ref{Delta_m02}). At the critical point, such surface Dirac points emerge from the bulk Dirac points, and away from the transition, they approach and eventually merge with the Dirac point at the BZ center (see Fig. \ref{Delta_m08}). We note that these Dirac points exist solely due to the presence of mirror symmetry: the latter is essential to consider states with different mirror eigenvalues separately, and thus to conclude that their 'handedness' changes at different sides of the transition.

We also note that, since Dirac metal phase is a critical point, where the surface state branches change their 'handedness', or, in other words, they change the bands, at which they terminate, we can conclude that, in Dirac metal, surface states terminate exactly at the gap-closing points, i.e. at the bulk Dirac points (this is shown on Fig. \ref{Delta_0}). This fact makes the surface states in Dirac metal similar to Weyl metal.

\subsection{Broken mirror symmetry}
\label{BrokenMirrorSymmetry}
\begin{figure}[h]
\centering
	\begin{subfigure}[b]{0.24\textwidth}
		\includegraphics[width=4.8cm,angle=0]
		{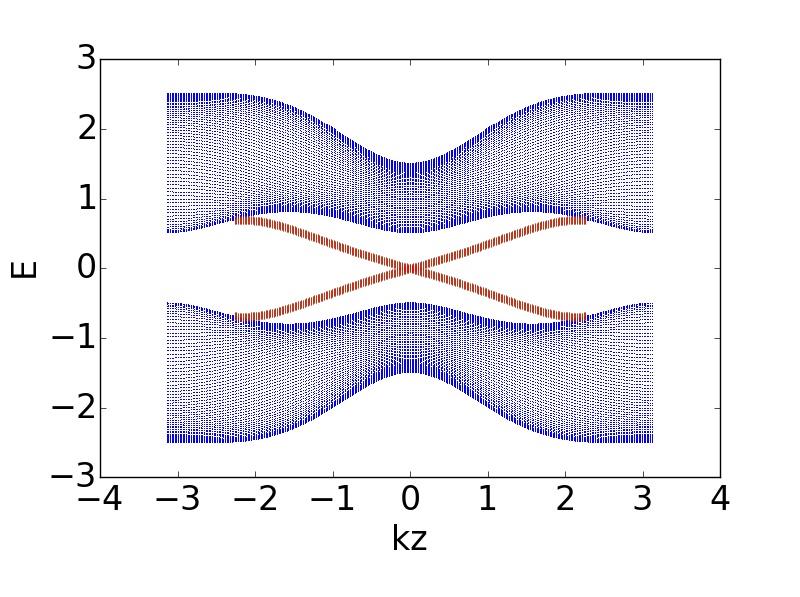}
		\vspace{-0.5cm}
		\subcaption{}	
		\label{Delta_08_MB}
	\end{subfigure}
	\begin{subfigure}[b]{0.23\textwidth}
		\includegraphics[width=4.8cm,angle=0]
		{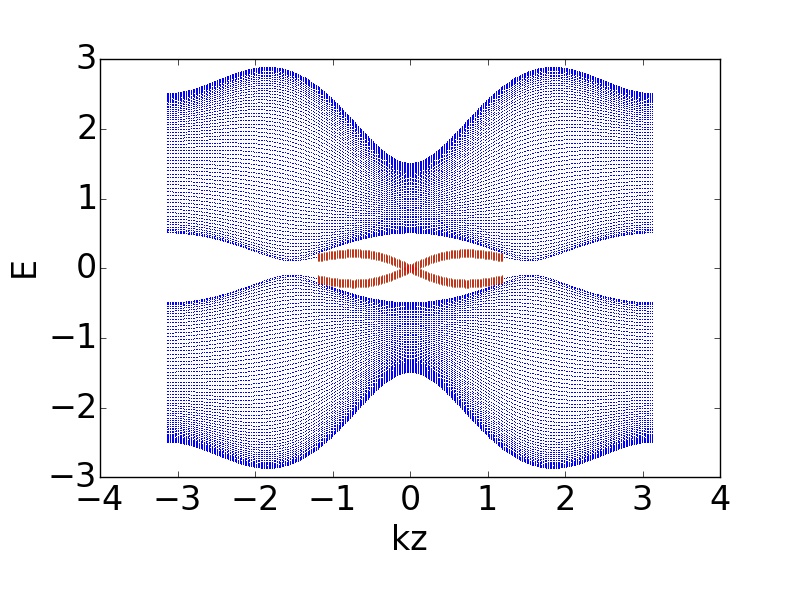}
		\vspace{-0.5cm}
		\subcaption{}
		\label{Delta_0_MB}
	\end{subfigure}
	\begin{subfigure}[b]{0.24\textwidth}
		\includegraphics[width=4.8cm,angle=0]
		{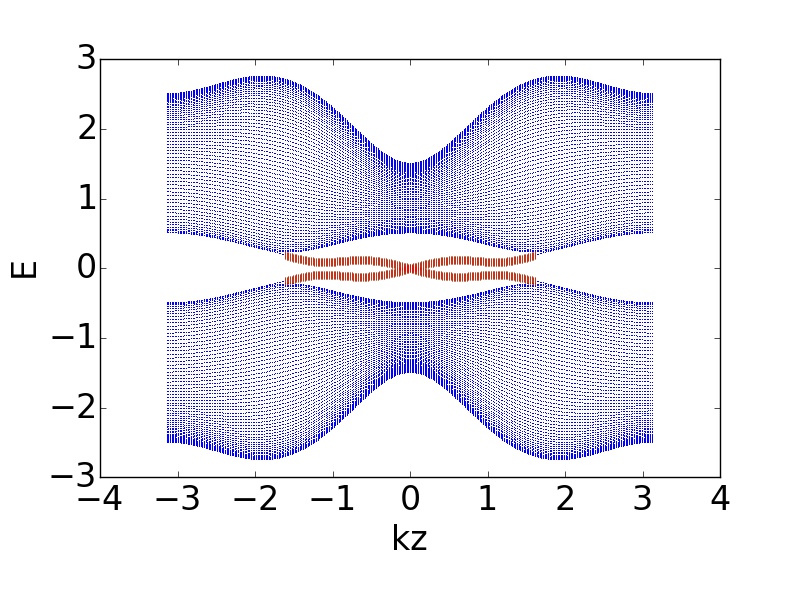}
		\vspace{-0.5cm}
		\subcaption{}
		\label{Delta_m02_MB}
	\end{subfigure}
	\begin{subfigure}[b]{0.23\textwidth}
		\includegraphics[width=4.8cm,angle=0]
		{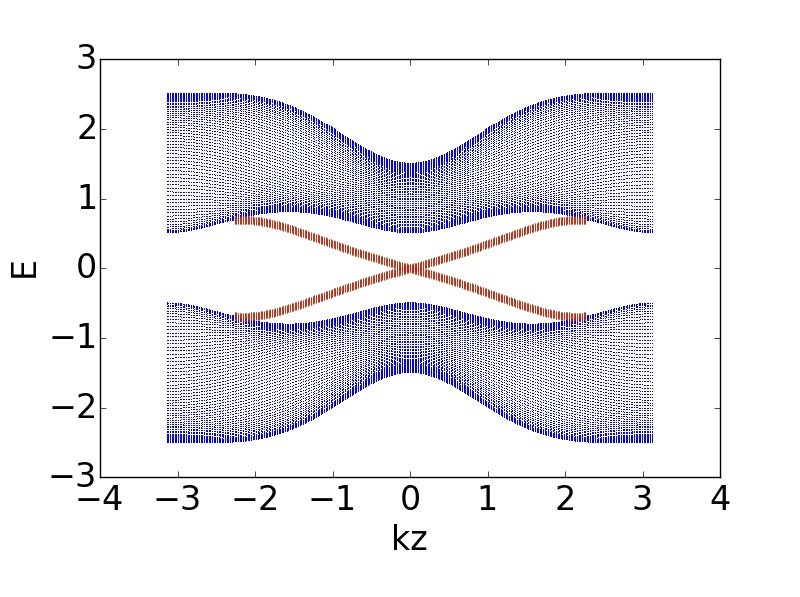}
		\vspace{-0.5cm}
		\subcaption{}
		\label{Delta_m08_MB}
	\end{subfigure}
\caption{Dispersion strucure of the model (\ref{HFixedMirror}) in the case of broken mirror symmetry. The values of $\Delta_1$ are $0.8$ (\subref{Delta_08}), $0.0$ (\subref{Delta_0}), $-0.2$ (\subref{Delta_m02}), $-0.8$ (\subref{Delta_m08}). The parameters are: $v_F=1.0$, $m=0.5$, $b_{xy}=0.5$, $b_z = 0.5$, $\gamma=1.0$, $\Delta_2 = 0.1$.   }
\label{SurfaceStatesMirrorBroken}
\end{figure}

Once we established, that mirror symmetry makes it possible to view Dirac metal as a critical point between two different topological phases, we would like to ask if the presence of the mirror symmetry is necessary. To explore the behavior of the insulating phase, we add to the model defined by the Eqs. (\ref{DiracHam}, \ref{Gamma4Term}) an additional mirror symmetry breaking term:
\begin{eqnarray}
H_2 = \Delta_2 \sin k_z \Gamma_5.
\label{MirrorBreakingTerm}
\end{eqnarray}
This term does not commute with the mirror symmetry operator over $xz$ plane, and therefore, the bands of the full Hamiltonian at $k_y=0$ can no longer be characterized by their mirror eigenvalues. Indeed, by continuous varying of the parameters $(\Delta_1 , \Delta_2)$, it is possible to smoothly transform the system between the two different phases. We demonstrate the resulting energy spectrum on the Fig. \ref{SurfaceStatesMirrorBroken}, where we take the state at $\Delta_1 >0$, introduce non-zero $\Delta_2$, then smoothly vary $\Delta_1$ from positive to negative value, and eventually 'turn off ' $\Delta_2$. The system undergoes such transformation smoothly, i.e. without gap closing. In other words, two phases, which are topologically distinct in the presence of the mirror symmetry, are not distinct if the mirror symmetry is broken.

Since the system with broken mirror symmetry still possesses $\mathtt{Z}_2$ invariant, its surface states spectrum contains the Dirac point in the center of the BZ. However, since the surface states are no more protected by the mirror symmetry, the other two Dirac points emerging at $k_z \ne 0$ (which are shown on the Fig. \ref{Delta_m02}) become gapped. 

So far, we have found that the presence of mirror symmetry is necessary to say, that Dirac metal is a critical point between two different topological phases. One may think, that only in the presence of mirror symmetry, we can conclude that surface states in Dirac metal terminate exactly at the Dirac points. However, we can look at the problem in a different way. We have written a possible term (\ref{MirrorBreakingTerm}) that breaks the mirror symmetry, and together with it, breaks rotational symmetry, thus gapping the Dirac points out. Now, let us ask: can we add to our model a term that would break the mirror symmetry, but would not gap the bulk Dirac points? Evidently, since the Dirac points are protected by the rotational, time-reversal and inversion symmetries, such term has to break the mirror symmetry, but at the same time, not to break any of the time-reversal, inversion, or rotaional symmetries. The question about all possible terms satisfying the latter symmetries was studied in details in the Ref. \cite{ NComms5.5898}, and it was found that they are strongly constrained. Indeed, it was found that, in the presence of TR, inversion and rotational symmetries, Hamiltonians can take just a few possible forms, and all their low energy expansions were written explicitly. The model considered in this paper is one of these explicit examples, and thus if we add to it a term breaking the mirror symmetry, but not breaking the other symmetries, it has to remain in that class. Therefore, we can conclude that if such a term exists, it has to be subleading in the Hamiltonian (\ref{DiracHam}), and thus negligible at small transverse momenta $k_{x, y}$. 

 However, near the Dirac points, the surface states have very large penetration depth, which corresponds to very small imaginary momentum. As a result, near the Dirac points, contribution to the surface states from all subleading terms including mirror symmetry breaking terms can be neglected. The surface Dirac cone at $k_z=0$ is protected by $\mathtt{Z}_2$ invariant, and therefore also persist if the mirror symmetry is broken. Thus we can conclude that the main result of our paper, i.e. the fact that surface states in Dirac metal terminate at Dirac points remains, if mirror symmetry is slightly broken. However its presence makes it possible to compute them analytically and explore their evolution after gap opening. 

\section{Discussion}
\label{Conclusions}

In this paper, we have established that Dirac metal with two spatially separated Dirac points can be viewed as a critical point between two different topological phases. In the case of non-unity inversion matrix, the transition is between two $\mathtt{Z}_2$ insulators with opposite mirror Chern numbers, defined in a plane passing through the Dirac points. Thus, Dirac metal can be viewed as a critical point between two phases of topological crystalline insulators.

We have studied the resulting surface states spectrum of the system. We found that in Dirac metal, surface states always terminate at the Dirac points. Between them, the surface states have Dirac cone at the BZ center, but are gapped everywhere else. Thus, our results can be viewed as a complement to the Ref. \cite{Kargarian02082016}, where it was also claimed, that surface states in Dirac metal are, generally, gapped.

 It is interesting to point out, that Fermi arcs in Weyl and Dirac metals can be probed in quantum oscillation experiments \cite{NatCommun5.6161}. Indeed, such experiment in Dirac metal $\mathrm{Cd_3 As_2}$ has been realized in the work \cite{Moll2016}, and later in Ref. \cite{Zhang2017}, where its dependence on Fermi energy was explored. Our results, consistently with the Ref. \cite{Kargarian02082016}, predict that electrons can form closed cyclotron orbits connecting the Dirac points only above certain Fermi energy, when the surface states connect two opposite Dirac cones (see Fig. \ref{E025}). However, in Ref. \cite{Zhang2017}, transition between two regimes at small and large Fermi energies (Fig. \ref{E015} vs \ref{E025}), has not been detected. We think that, there is no contradiction between the experiment and predictions. Indeed, we have estimated the gap between the Fermi arcs in Dirac metal to be proportional to small parameter $\gamma$, which effects have never been measured experimentally. Thus, we think that the value of $\gamma$ in Dirac metal $\mathrm{Cd_3 As_2}$ is accidentally very small to be observed in the past experiments. We suggest, that in the future experiments, it might be a good idea to measure quantum oscillation dependence on chemical potential more accurately, and perhaps in different materials.

We have found, that when, after adding rotation symmetry breaking terms, the system becomes a topological crystalline insulator, its surface states always form Dirac point in the BZ center protected by $\mathtt{Z}_2$ invariant. However, the surface bands may also have two other Dirac points protected solely by the mirror symmetry. We also would like to note that, in the Ref.  \cite{NComms5.5898} it was predicted another kind of type \Roman{TypeOne} Dirac metal with unity inversion matrix, which to best of our knowledge, have not yet been realized experimentally. In contrast to our model, it does not possess $\mathtt{Z}_2$ invariant, and therefore its surface states spectrum will not have Dirac point at the BZ center. However, such model still possesses the mirror symmetry, and thus it can also be viewed as a critical point between two topologically distinct phases. We checked that its mirror Chern number changes from $0$ to $2$. Similarly, in the topological phase, it may possess surface Dirac points protected by the mirror symmetry.

Finally, we would like to discuss possible realizations of the described topological phase transition. Various scenarios of mirror symmetry breaking were proposed in the general context of topological crystalline insulators (see Ref. \cite{doi:10.1146/annurev-conmatphys-031214-014501} for review), and a few of them can be used to realize rotational symmetry breaking as well. For example, rotational symmetry in Dirac metal can be directly broken by structural phase transition, which in turn can be created either by change of temperature \cite{PhysRevB.91.214517}, or by applying high pressure\cite{SciRep6.24137, PhysRevB.93.045118, PhysRevB.96.075112, C4CP05115G}. It has been verified experimentally that Dirac metal can be converted into topological insulator, but it has not yet been verified that Dirac metal can be turned into topological crystalline insulator.

Another possible way to find topological phase transition in Dirac metal is to study spontaneous symmetry breaking, which can occur through interactions. In particular, it was predicted that at sufficiently large electron-electron interaction, rotational symmetry can be broken by nematic order \cite{PhysRevB.93.041108}. In addition, rotational symmetry was predicted to be broken in the presence of superconductivity \cite{PhysRevB.95.201110}.

\begin{acknowledgements} 
The author would like to thank professor Anton Burkov for numerous discussions. Financial support was provided by NSERC of Canada. 
\end{acknowledgements}



\appendix

\section{Analytical calculation of Chern number}
\label{CN}
In this section, we present the explicit calculation of Chern number in a model (\ref{HFixedMirror}) of the main text. We are doing it, because, despite the fact that this type of Hamiltonians is widely studied (see e.g. Ref. \cite{10.2307/j.ctt19cc2gc} ), we are not familiar with any references, where the analytical calculation has been presented.

Since the Chern number is integer only if the underlying space is compact, it is essential that our model is placed on the lattice. For simplicity, we assume that the Hamiltonian has a general form: 
\begin{eqnarray}
H =  d_a (\vec k)  \tau_a,
\label{Gen_Ham}
\end{eqnarray}
where, in our case: 
\begin{eqnarray}
d_1 &=& v_F \sin k_x,  
\label{d_i_lattice}  \\
d_2  &=& \sin k_z \left( \gamma (1-\cos k_x)  + \Delta_1 \right),  
\nonumber\\
d_3 &=& \pm \left[  m - 2 b_{xy} (1 - \cos k_x) - b_z(1 - \cos k_z ) \right].
\nonumber
\end{eqnarray}
We assume that $v_F >0$, but $\Delta_1$ may have any sign. We also assume that the parameters $m, b_{xy}, b_z$ are chosen in such way that, at $\Delta_1 = 0$, the system possesses Dirac points, i.e. $m(k)$ changes sign within the BZ, and that, at $k_z=0$, there are topological surface states in $xy$ plane. Strinctly speaking, details of the further derivation may depend on the numerical values of the parameters, but since the method is similar, we assume that our parameters are the same as on the Fig. \ref{SurfaceStates}, i.e. 
\begin{eqnarray}
 m &=& 0.5,
\nonumber\\
 b_{xy} &=& 0.5, 
\nonumber\\
b_z &=& 0.5.
\label{ParameterValues}
\end{eqnarray}

The expression for Berry connection in the model (\ref{Gen_Ham}) is well-known:
\begin{eqnarray}
\mathcal{A}_i = \frac{d_1 \partial_i d_2  -  d_2 \partial_i d_1}{2d(d+d_3)},
\nonumber
\end{eqnarray}
and Chern number is defined as its integral over the boundary of the Brillouin zone:
\begin{eqnarray}
C = \frac{1}{2\pi} \int \vec dk \vec{ \mathcal{A}}.
\label{Chern_Num_Def} 
\end{eqnarray}

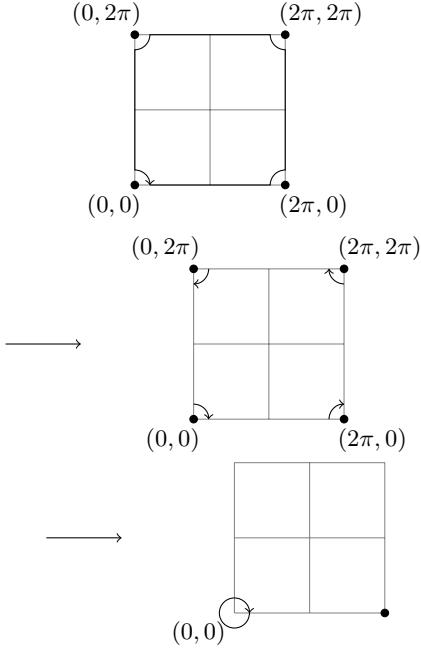
\begin{figure}[h]
\begin{tikzpicture}
\draw[help lines] (0,0) grid (2,2);
\draw[->] (0.2, 0) -- (1.8,0) arc(180: 90: 0.2) -- (2, 1.8)  arc(270: 180: 0.2) -- (0.2,2) arc(0:-90: 0.2) -- (0,0.2) arc(90:0:0.2);
\draw[fill] (0,0) circle [radius=0.05];
\node[below left] at (0.2,0) {$(0, 0)$};
\draw[fill] (2,0) circle [radius=0.05];
\node[below right] at (1.8,0) {$(2\pi, 0)$};
\draw[fill] (2,2) circle [radius=0.05];
\node[above right] at (1.8, 2) {$(2\pi, 2\pi)$};
\draw[fill] (0,2) circle [radius=0.05];
\node[above left] at (0.2, 2) {$(0, 2\pi)$};
\end{tikzpicture}
\begin{tikzpicture}
\draw[help lines] (6,0) grid (8,2);,
\draw[->] (3.5, 1) -- (4.5, 1);
\draw[->]  (7.8,0) arc(180: 90: 0.2);
\draw[->]  (8, 1.8)  arc(270: 180: 0.2); 
\draw[->]  (6.2,2) arc(0:-90: 0.2); 
\draw[->]  (6,0.2) arc(90:0:0.2);
\draw[fill] (6,0) circle [radius=0.05];
\draw[fill] (8,0) circle [radius=0.05];
\draw[fill] (8,2) circle [radius=0.05];
\draw[fill] (6,2) circle [radius=0.05];
\node[below left] at (6.2,0) {$(0, 0)$};
\node[below right] at (7.8,0) {$(2\pi, 0)$};
\node[above right] at (7.8, 2) {$(2\pi, 2\pi)$};
\node[above left] at (6.2, 2) {$(0, 2\pi)$};
\end{tikzpicture}
\begin{tikzpicture}
\draw[help lines] (12,0) grid (14,2);,
\draw[->] (9.5, 1) -- (10.5, 1);
\draw[->] (12.2,0) arc(0: -360: 0.2);
\draw[fill] (14, 0) circle [radius = 0.05];
\node [below left] at (12, 0) {$(0, 0)$};
\end{tikzpicture}
\caption{Integration contour for $\vec{\mathcal{A}}$ in the plane $(k_x, k_z)$ in the model defined by $H_{+}$  (Eq. \ref{HFixedMirror}). }
\label{ContourIntegration0_2pi}
\end{figure}

Since the BZ is periodic, the integral is, in fact, taken over a closed contour passing along the lines, where $k_x$ or $k_y$ are equal to $0$ or $2\pi$. If $\vec{\mathcal{A}}$ were smoothly and uniquely defined over the whole BZ, the integral over the line $k_{x,y} = 0$ would cancel out the integral over the line $k_{x,y} = 2\pi$, and, as a result, the Chern number would be zero. However, in our case, there are several points, where the Berry connection is not well-defined, and therefore the integration contour has to be deformed to go around such points. More particularly, the Berry connection diverges if $d=0$ or $d+d_3 = 0$. The first condition is not satisfied at general values of $m$, $b_{xy}$, $b_z$, but the second is satisfied, when $d_1 = d_2 = 0$, and $d_3 < 0$. If we consider the case of positive sign in the expression for $d_3$ (Eq. \ref{d_i_lattice}) and assume that numerical values of our parameters are as in the  Eq. (\ref{ParameterValues}), the Berry connection diverges at $k_{x, z} = 0, 2\pi$. Thus the actual integration contour looks as
shown on the Fig. \ref{ContourIntegration0_2pi}.

On that figure, we have shown the way, in which the integration is simplified: integration along constant $k_{x, z} = 0$ cancels out integration along $k_{x, z} = 2\pi$. Thus, integration along the whole contour is equivalent to integration along four quarter-circles surrounding the poles of $\mathcal{\vec{A}}$, which is, in turn, equivalent to integration along one small circle surrounding the point $(k_x, k_z) = (0, 0)$. The latter is performed straightforwardly in polar coordinates in $(k_z, k_z)$ plane:
\begin{eqnarray}
v_F k_x = k \cos \varphi,
\nonumber\\
|\Delta| k_z = k \sin \varphi.
\nonumber
\end{eqnarray}
Since $k_{x,z}$ within the integration contour is small, we can expand $d_a$ up to linear order in $k_{x, z}$, and assume $d_{1, 2} < < d_3$, thus obtaining:
\begin{eqnarray}
\vec{\mathcal{A}} d\vec{k} = \mathrm{sign}\Delta_1 d\varphi.
\nonumber
\end{eqnarray}
This results the Chern number being equal to $C = \mathrm{sign} \Delta_1$. 

\begin{figure}[h]
\begin{tikzpicture}
\draw[help lines, thin] (0,0) grid (2,2);
\draw[->] (0, 0) -- (0.8,0) arc(180: 0: 0.2)-- (2, 0) -- (2, 0.8)arc(270: 90: 0.2) -- (2, 2)  -- (1.2, 2) arc(0:-180: 0.2) -- (0, 2)-- (0, 1.2) arc(90:-90:0.2) -- (0, 0);
\draw[fill, thick] (1,0) circle [radius=0.05];
\node[below] at (1, 0.0) {$(\pi, 0)$};
\draw[fill] (2,1) circle [radius=0.05];
\node[right] at (2, 1) {$(2\pi, \pi)$};
\draw[fill] (1,2) circle [radius=0.05];
\node[above] at (1, 1.9) {$(\pi, 2\pi)$};
\draw[fill] (0,1) circle [radius=0.05];
\node[left] at (0, 1) {$(0, \pi)$};
\draw[->]  (1.2, 1) arc(360:0:0.2);
\draw[fill] (1, 1) circle [radius=0.05];
\node[above] at (1, 1.1) {$(\pi, \pi)$};
\end{tikzpicture}
\begin{tikzpicture}
\draw[->] (3.5, 1) -- (4.5, 1);
\draw[help lines] (6,0) grid (8,2);
\draw[->]  (6, 1.2) arc(90: -90: 0.2);
\draw[->]  (6.8, 0)  arc(180: 0: 0.2); 
\draw[->]  (8, 0.8) arc(270: 90: 0.2); 
\draw[->]  (7.2, 2) arc(0:-180:0.2);
\draw[fill] (7, 0) circle [radius=0.05];
\draw[fill] (8, 1) circle [radius=0.05];
\draw[fill] (7, 2) circle [radius=0.05];
\draw[fill] (6, 1) circle [radius=0.05];
\draw[->]  (7.2, 1) arc(360:0:0.2);
\draw[fill] (7, 1) circle [radius=0.05];
\node[below] at (7, 0.0) {$(\pi, 0)$};
\node[right] at (8, 1) {$(2\pi, \pi)$};
\node[above] at (7, 1.1) {$(\pi, \pi)$};
\node[above] at (7, 1.9) {$(\pi, 2\pi)$};
\node[left] at (6, 1) {$(0, \pi)$};
\end{tikzpicture}
\begin{tikzpicture}
\draw[help lines] (12,0) grid (14,2);
\draw[->] (9.5, 1) -- (10.5, 1);
\draw[->] (12.2,1) arc(0: -360: 0.2);
\draw[fill] (12, 1) circle [radius=0.05];
\draw[->] (13.2,0) arc(0: -360: 0.2);
\draw[fill] (13, 0) circle [radius = 0.05];
\draw[->] (13.2,1) arc(0: -360: 0.2);
\draw[fill] (13, 1) circle [radius = 0.05];
\end{tikzpicture}
\caption{Integration contour for $\vec{\mathcal{A}}$ in the model defined by $H_{-}$  (Eq. \ref{HFixedMirror}). }
\label{ContourIntegration_pi_pi}
\end{figure}
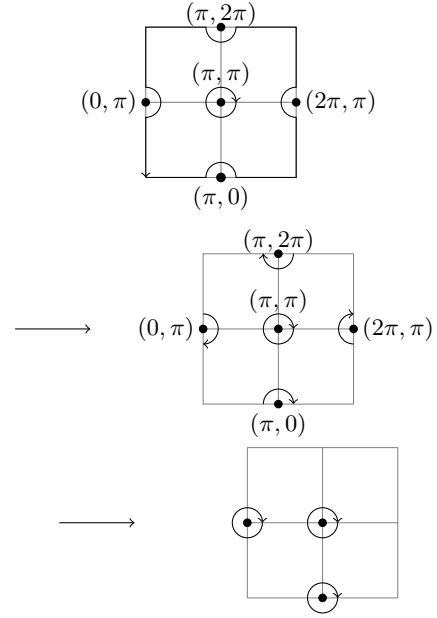

In the case of negative sign in the Eq. (\ref{d_i_lattice}), the reasoning is similar. The only difference is that $\vec{\mathcal{A}}$ diverges at the points $(k_x, k_z) = (0, \pi), (\pi, 2\pi), (2\pi, \pi), (\pi, 0), (\pi, \pi)$, and thus the integration contour has the form, shown on the Fig. \ref{ContourIntegration_pi_pi}.

In this case, the integration contour is simplified in a similar way leading to three small contours surrounding the points $(0, \pi)$, $(\pi, 0)$, $(\pi, \pi)$, which are, in turn, integrated in exactly same way, as in the previous case. The resulting answer is $C = - \mathrm{sign} \Delta_1$. 

Thus, the final answer for the Chern number in the model (\ref{d_i_lattice}): $C = \pm \mathrm{sign} \Delta_1$.



\section{Analytical solution for surface states}
\label{SecSurfaceStates}

In this section, we present analytical solution for the surface states in the model determined by the total Hamiltonian $H = H_0 + H_1$, where $H_{0, 1}$ are defined according to the Eqs. (\ref{DiracHam}, \ref{Gamma4Term}). As we mentioned previously, we are interested in the special case $k_y=0$, when the total Hamiltonian decouples into two Hamiltonians (\ref{HFixedMirror}). We use method, similar to the Refs. \cite{RevModPhys.83.1057} \cite{PhysRevB.82.045122}.

We are interested in the solutions obeying Dirichlet boundary conditions at the sample boundary, i.e. at $x=0$, and exponentially decaying away from the surface.  To satisfy both these conditions, we take an ansatz:
\begin{eqnarray}
\Psi(x, k_z)  = \left( e^{-\lambda_1 x} - e^{-\lambda_2 x}  \right) \psi(k_z).
\nonumber
\end{eqnarray}
This equation implies that the state $\psi$ has to be an eigenstate of the Hamiltonian at two different values of $\lambda$, i.e.
\begin{eqnarray}
E_{\pm} \psi_{\pm} = H_{\pm}(-i\lambda_1, k_z) \psi = H_{\pm}(-i\lambda_2, k_z) \psi_{\pm},
\label{SurfaceEignevector}
\end{eqnarray}
which, in turn, means that $\psi$ has to be an eigenstate with zero eigenvalue of the difference between the Hamiltonians at $\lambda_{1, 2}$:
\begin{eqnarray}
0 = \frac{\left[ H_{\pm}(-i\lambda_1, k_z) - H_{\pm}(-i\lambda_2, k_z) \right] \psi }{\lambda_1 - \lambda_2}.
\label{DiffHamEq}
\end{eqnarray}
The fact that the operator, entering the last equation, has zero eigenvalue, implies that $\mathrm{det} \left( H_{\pm}(-i\lambda_1) - H_{\pm}(-i\lambda_2) \right) = 0$, which, in turn, leads to the condition:
\begin{eqnarray}
| \lambda_1 + \lambda_2 | = \frac{v_F}{\sqrt{b_{xy}^2 + \left( \frac{\gamma k_z}{2}\right)^2}}.
\label{LambdaSum}
\end{eqnarray}
Here, the case of positive (negative) $\lambda_1 + \lambda_2$ corresponds to the state decaying at $x>0$ ($x<0$). We focus our reasoning on the case of the positive sum. An eigenvector satisfying the Eqs. (\ref{DiffHamEq}) and (\ref{LambdaSum}) has the form:
\begin{eqnarray}
\psi = 
	\left(
	\begin{array}{c}
		\sqrt{b_{xy}^2 +\left( \frac{\gamma k_z}{2}\right)^2 } - \frac{\gamma k_z}{2}
		\\
		\mp ib_{xy}
	\end{array}
	\right)
\label{psi_expr}
\end{eqnarray}

The last equation is the answer for an eigenvector of the surface state, but we still have to find its energy and penetration depth. To do it, we substitute this vector (\ref{psi_expr}) into the Eq.  (\ref{SurfaceEignevector}) and require it to be the actual eigenvector. Such condition results, that  $\lambda_{1, 2}$ has to be equal to:
\begin{eqnarray}
\lambda_{1, 2} = \frac{v_F  \pm  \sqrt{v_F^2 - 4 b_{xy}(m - \frac{mb_z^2}{2}) + 2 \Delta_1 \gamma k_z^2  }}{2 \sqrt{b_{xy}^2 + \left( \frac{\gamma k_z}{2} \right)^2}}.
\label{LambdaAnswer}
\end{eqnarray}
If we substitute it into the Hamiltonian, we can find answer for the energy:
\begin{eqnarray}
E = \mp \frac{\gamma k_z}{2 \sqrt{b_{xy}^2 + \frac{\gamma^2 k_z^2}{4}  }}  
\left( m - \frac{b_z k_z^2}{2}  + \frac{2 \Delta_1 b_{xy}}{\gamma} \right)
\label{EnergyAnswer}
\end{eqnarray}

Thus, we obtain that the surface spectrum consists of two braches, corresponding to two different mirror eigenvalues.  One can check that the Eq. (\ref{EnergyAnswer}) satisfies all properties listed in the main text: two branches cross at the center of the BZ, become gapped away from it, and, in the case of zero bulk gap $\Delta_1$, they cross again at the bulk Dirac points, i.e. when $m - \frac{b_z k_z^2}{2} = 0$. One can also check that at these points, the penetration depth (\ref{LambdaAnswer}) becomes infinite. Finally, by analyzing the behavior of the functions (\ref{EnergyAnswer}, \ref{LambdaAnswer}), one can check explicitly that at non-zero gap $\Delta_1$, the surfaces states change their handedness, if $\Delta_1$ changes sign.

\bibliographystyle{apsrev4-1}
\bibliography{Bib_Dirac}

\end{document}